\documentstyle[twoside,fleqn,espcrc2]{article}

\def\beqa{\begin{eqnarray}}
\def\eeqa{\end{eqnarray}}

\newcommand{\AmS}{{\protect\the\textfont2
  A\kern-.1667em\lower.5ex\hbox{M}\kern-.125emS}}

\title{Perturbative gravitational couplings and Siegel modular forms in 
$D=4,N=2$ string models}

\author{G. L. Cardoso,\address{Theory Division, CERN,
        CH-1211 Geneva 23, Switzerland} %
         \thanks{Contribution to the Proceedings of 
the 30th International Symposium Ahrenshoop
on the Theory of Elementary Particles, Buckow, August 27-31, 1996}}

\begin{document}

\begin{abstract}

We consider 
four-parameter $D=4,N=2$ string models with Hodge numbers
$(4,214 - 12n)$ and $(4,148)$, and
we express 
their perturbative Wilsonian gravitational coupling $F_1$ 
in terms of Siegel modular forms.

\end{abstract}

\maketitle

\section{Introduction}

String-string dualities 
between heterotic strings on $K3 \times T_2$ and corresponding
type II strings on suitably chosen Calabi--Yau threefolds 
in four space-time
dimensions have been successfully tested in \cite{KV}--\cite{HM2}.

One interesting aspect in the study of these $N=2$ string-string dualities
in four space-time
dimensions is the appearence \cite{HM} of certain modular forms
in the low-energy effective action of these theories.  Here, we will
focus on $D=4, N=2$ four-parameter string models with Hodge numbers
$(4,214 - 12n)$ and $(4,148)$, and we will express their perturbative
Wilsonian gravitational coupling $F_1$ in terms of Siegel modular forms.
For the models with Hodge numbers $(4,214-12n)$, this has been
achieved in \cite{CCL}, and the results presented below for these models
represent a short summary of the ones contained in \cite{CCL}.

\section{$D=4, N=2$ four-parameter string \\ models}

The four-parameter models we will be considering in the
following have a perturbative heterotic description in terms of 
compactifications of the 
 $E_8^{(1)} \times
E_8^{(2)}$ heterotic string on $K3 \times T_2$.  They also have
a dual type II description in terms of compactifications of the
type II string on Calabi--Yau threefolds which are $K3$ fibrations
\cite{KLM,AL}.

We will consider two classes of models.  The first class of models can 
be obtained,
in the type II description, by compactifying on Calabi--Yau threefolds
with Hodge numbers $(h^{1,1},h^{2,1})=(4, 214-12n)$ and, consequently,
 Euler number
$\chi =  24n - 420$.  We will restrict the discussion to $n=0,1,2$.
These models have a perturbative heterotic description in terms
of compactifications of the heterotic $E_8^{(1)} \times
E_8^{(2)}$ string on $K3 \times T_2$ with $SU(2)$ gauge bundles
on the $K3$ with instanton numbers $(d_1,d_2)=(12-n, 12+n)$ \cite{KV,CF,AFIQ}.
In the heterotic description, the four moduli comprise the dilaton $S$,
the two toridal moduli $T$ and $U$ as well as a Wilson line modulus $V$.
The massless spectrum thus contains $N_V=5$ vector multiplets (here we
also count the
graviphoton) as well as $N_H=215-12n$ neutral hyper multiplets.  At the
transition surface $V=0$, the $U(1)_V$ associated with the Wilson line
modulus becomes enhanced to an $SU(2)$.  Let $N_V'=2$ and $N_H'=12n + 32$
denote the additional vector and hyper multiplets becoming massless
at $V=0$.  Then, as the $SU(2)$ gauge bosons get swallowed,
the transition to an $S$-$T$-$U$ model with $N_V=4$ vector and $N_H=244$
neutral hyper multiplets takes place \cite{BKKM}.

The second class of models can be obtained, in 
the type II description, 
by compactifying on Calabi--Yau threefolds with Hodge numbers 
$(h^{1,1},h^{2,1})=(4, 148)$ and, consequently, 
Euler number $\chi = - 288$.  These
Calabi--Yau threefolds are actually also elliptic, that is, they
are elliptically fibred with base ${\bf F}_n$ ($n=0,1,2$), and
their elliptic fibre is of the $E_7$ type \cite{AFIU,KMV}.
These models
also have a perturbative heterotic description,
apparently in terms of compactifications of the heterotic $E_8^{(1)}
\times E_8^{(2)}$ string on $K3 \times T_2$ with $U(1)$ backgrounds
on $K3$ \cite{GSW,AFIU}.  Again, the heterotic moduli will comprise the
dilaton $S$, the two toroidal moduli $T$ and $U$ as well as a 
fourth modulus $V$.  The massless spectrum thus contains, for $V \neq 0$,
$N_V=5$ vector multiplets as well as $N_H= 149$ neutral hyper multiplets.
There are also a total of 96 hyper multiplets charged under the $U(1)_V$
associated with the modulus $V$ \cite{GSW,AFIU}.  They are massive
at generic points in the moduli space ($V \neq 0$).  At $V=0$,
there is no additional gauge symmetry enhancement.  Instead, these 96
hyper multiplets become massless.  The $U(1)_V$ gauge boson is swallowed,
and again the transition to an $S$-$T$-$U$ model takes place ($\chi = -288
- 2 \times 96 = - 480$).

We will, in  the following, 
collectively denote all these four-parameter models
by $S$-$T$-$U$-$V$ models.

\section{The perturbative prepotential $F$}

In this section we briefly discuss the 
relation
between the type II and the heterotic prepotentials for the 
$S$-$T$-$U$-$V$ models.  It will turn out to be useful to introduce
a parameter $t$ in order to describe the perturbative corrections
to the heterotic prepotential for the models described aboved.
For the models with Hodge numbers $(4,214-12n)$,
the parameter $t$ will take the value $t=1$,
whereas for the models with Hodge numbers $(4,148)$
it will take the value $t=2$.

In the following, the dilaton $S$ is defined to be 
$S = \frac{4 \pi}{g^2} - i \frac{\theta}{2 \pi}$.

For an $S$-$T$-$U$-$V$ model,
the perturbative
heterotic prepotential is given by
\beqa
F^{\rm het} 
= - S (TU - t V^2) + h(T,U,V) \;\;,\;\;\;\;\;\;\;\;\;\;\;\;\;\;\;\;
\label{prepstuv} \nonumber\\
h(T,U,V) =  p(T,U,V) -\frac{c(0)\zeta(3)}{8 \pi^3} \;\;\;\;\;\;\;\;\;\;\;\;
\;\;\;\;\;\;\;\;\;
\\
-\frac{1}{4\pi^3}\sum_{k,l,b\in {\bf Z} \atop
(k,l,b)>0} c(4tkl-b^2) Li_3(e^{-2 \pi(kT+lU+bV)}). \nonumber
\eeqa
The
condition
$(k,l,b)>0$ means that \cite{HM}: either $k > 0, l , b \in {\bf Z}$ or 
$k=0,l>0, b \in {\bf Z}$ 
or $k=l=0, b > 0$.

The term $ - S(TU-tV^2)$ denotes the tree-level
prepotential associated with the classical vector moduli space
$\frac{SU(1,1)}{U(1)} \times
{\cal H}^+_t$, where ${\cal H}^+_t$ 
has a representation as a three-dimensional
tube domain of type IV \cite{gritnik3}.   ${\cal H}^+_t$ is isomorphic
to the Siegel upper half plane ${\cal H}_2$ of complex dimension 3.

The term $p(T,U,V)$ denotes a cubic polynomial, which is generated at
one-loop, and 
which depends on the particular instanton embedding.

The coefficients $c(4tkl-b^2)$ appearing
in the prepotential are, presumably, 
the expansion coefficients of a Jacobi form
$f(\tau,z)$ of weight $-2$ and index $t$
\beqa
f(\tau, z) = \sum_{n\geq-1, l \in {\bf Z}}
c(4 t m - l^2) q^m r^l \;\;,
\eeqa
where $q=\exp{2 \pi i \tau}, r = \exp{2 \pi i z}$.
For the case $t=1$, the appropriate Jacobi form of weight $-2$
and index $1$ is given by \cite{K2,CCL}
\beqa
f(\tau, z) = \frac{1}{\eta^{24}} \left(\frac{12-n}{24} E_6 E_{4,1} + 
\frac{12+n}{24} E_4 E_{6,1} \right).
\eeqa
It would be interesting to find
the appropriate Jacobi form for the $t=2$ case.

It can be checked, by comparison with the corresponding type II
prepotential, that the expansion coefficients $c(4tkl-b^2)$ appearing
in (\ref{prepstuv}) indeed only depend on the combination
$4 tkl -b^2$.  In order to do so, one has to consider 
the contributions of the world sheet 
instantons to the type II prepotential which, for 
four-parameter models, are 
genericallly given by 
\beqa
F^{II}_{\rm inst} =-{1\over ( 2\pi)^3}\sum_{d_1, d_2, d_3, d_4}
n^r_{d_1,d_2, d_3,
d_4}Li_3(\prod_{i=1}^4q_i^{d_i}),
\label{instanton}
\eeqa
where $q_i = \exp{(-2 \pi t_i)}$.  The
$n^r_{d_1,d_2,d_3,d_4}$ denote the rational instanton numbers.
In order to match (\ref{instanton}) with (\ref{prepstuv}), we 
will perform the following identification of the type II K\"ahler
class moduli $t_i$ with the heterotic moduli $S$, $T$, $U$ and $V$ \cite{CCL}
\beqa
t_1&=&U-2V,\qquad t_2= S-{n\over 2}T-(1-{n\over 2})U,\nonumber\\
t_3&=&T-U,\qquad t_4=V  \;\;,
\label{coordin}
\eeqa
which is valid in the chamber $T>U>2V$.
Then, the 
heterotic weak coupling limit $S \rightarrow \infty$ corresponds
to taking the large K\"ahler class limit $t_2 \rightarrow \infty$.  In this
limit, only the instanton numbers with $d_2=0$ contribute
in the above sum (\ref{instanton}).  Using the identification 
$kT+lU+bV=d_1t_1+d_3t_3+d_4t_4$, it follows that (independently of $n$)
\begin{eqnarray}
k&=&d_3 \;\;, \nonumber\\
l&=&d_1-d_3   \;\;,\nonumber\\
b&=&d_4-2d_1  \;\;.
\end{eqnarray}
Then, (\ref{instanton}) turns into
\beqa
F^{II}_{\rm inst}=-{1\over (2\pi)^3}\sum_{k,l,b}n^r_{k,l,b}
Li_3(e^{-2\pi(kT+lU+bV)}).
\label{largespr}
\eeqa
Comparison with (\ref{prepstuv}) shows that the rational instanton
numbers should be related to the expansion coefficients $c(4tkl-b^2)$
by
\beqa
n^r_{k,l,b} = - 2 c(4tkl-b^2) \;\;,
\label{insmod}
\eeqa
in which case they would have to satisfy the following non-trivial
constraint
\beqa
n^r_{k,l,b}=n^r(4tkl-b^2) \;\;.
\label{constraint}
\eeqa
It can now be checked that the constraint (\ref{constraint})
is indeed satisfied for the models discussed here.  This, in particular,
also holds for the model\footnote{
We would like to thank P. Mayr for providing us with the 
intersection numbers and the rational instanton numbers
 for the $X^{(4)}$ model in the $K3$ phase.} with Hodge numbers $(4,148)$
 based on the elliptically fibred Calabi--Yau
threefold with $E_7$
fibre ${\bf P}^{1,1,2}[4]$ and base ${\bf F}_1$
(denoted by $X^{(4)}$ in \cite{KMV}).  By inspection
of the rational instanton numbers for some of the four-parameter models under
consideration, it can also be inferred that 
\beqa
c(N) &=& 0 \;\;,\;\; N < - 4t \\
c(-4t)&=&1\;\;,\;\; 
c(-1) = - \frac{1}{2} N_H' \;\;,\;\; c(0) = \frac{1}{2} \chi
\;\;. \nonumber
\eeqa
For the $X^{(4)}$ model ($t=2$), for instance, one finds that
\beqa
c(-8) &=& 1 \;\;,\;\;
c(-4) = 0 \nonumber\\
c(-1) &=& - 48 \;\;,\;\; c(0) = -144 \;.
\eeqa
The truncation to a three-parameter
Calabi--Yau model is made by setting $t_4=0$.
The instanton numbers $n^{r}_{k,l}$ of the three-parameter
($S$-$T$-$U$) model are then 
given by \cite{BKKM}
\beqa
n^{r}_{k,l} = 
\sum_{b} n^r(4tkl-b^2) \;\;,\label{truncsum}
\eeqa
where the summation range over $b$ is finite.  
For example, when considering the transition from the 
four-parameter model $X^{(4)}$ to the three-parameter Calabi--Yau
model, 
$n^r_{0,1}= 96 + 288 + 96 = 480$.

The heterotic
cubic polynomial $p(T,U,V)$ can also 
be obtained from the cubic terms
of the associated type II polynomial, by making use of the map (\ref{coordin}).
Consider, for instance, the 
models with Hodge numbers $(4,214-12n)$, where $n=0,1,2$.
The cubic parts of
the type II prepotentials 
are given in \cite{BKKM,LSTY} and can be written as follows \cite{CCL}:
\beqa
F^{II}_{\rm cubic}&=&t_2(t_1^2+t_1t_3+4t_1t_4+2t_3t_4+3t_4^2)\nonumber \\
&+&{4\over 3}t_1^3+8t_1^2t_4+{n\over 2}t_1t_3^2
+(1+{n\over 2})t_1^2t_3 \nonumber\\
&+&2(n+2)t_1t_3t_4
+nt_3^2t_4
+(14-n)t_1t_4^2 \nonumber\\
&+& (4+n)t_3t_4^2+(8-n)t_4^3 . \label{cubicf}
\eeqa
Then, inserting (\ref{coordin}) into (\ref{cubicf}) yields
\beqa
F^{II}_{\rm cubic} &=& - F^{\rm het}_{\rm cubic} = S(TU-V^2)
+{1\over 3}U^3 \\
&+&({4\over 3}+n)V^3-(1+{n\over 2})UV^2-{n\over 2}
TV^2 \;\;. \nonumber
\label{cubicfa}
\eeqa
For $V=0$, $F^{\rm het}_{\rm cubic}$ 
precisely reduces to the cubic prepotential
of the $S$-$T$-$U$ models \cite{HKTY,LSTY}. 

Next, consider the models with Hodge numbers $(4,148)$.  For the
model with base ${\bf F}_1$, the cubic part of the type II prepotential
is, in the $K3$ phase ($c_2(J) = (92,24,36,88)$), given by
\beqa
\label{cubicx4}
F^{II}_{\rm cubic} &=& t_2(
t_1^2 + t_1 t_3 + 4 t_1 t_4 + 2 t_4^2  
+ 2 t_3 t_4) \nonumber\\
&+& \frac{4}{3} t_1^3 + 8  t_1^2 t_4 + 8 t_1 t_4^2 
+ \frac{8}{3} t_4^3 \nonumber\\
&+& \frac{3}{2}t_1^2 t_3+ 6 t_1 t_3 t_4
+ 3 t_3 t_4^2  + \frac{1}{2} t_1 t_3^2  \nonumber\\
&+& t_3^2 t_4 \;\;.
\eeqa
It can be checked that, for $t_2=0$, the above turns into the
prepotential of the three-parameter model with base ${\bf F}_1$
\cite{LSTY}.
Inserting (\ref{coordin}) (with $n=1$) into (\ref{cubicx4}) yields
\beqa
F^{II}_{\rm cubic} &=& - F^{\rm het}_{\rm cubic} = S(TU-2V^2)
+{1\over 3}U^3 \nonumber\\
&+& 8 V^3-4UV^2-2
TV^2 \;\;.
\eeqa

\section{BPS orbits}

An important role in the computation of the Wilsonian
gravitational coupling $F_1$ is played by BPS states \cite{FKLZ,VAFA,HM,CCLMR},
\beqa
F_1 \propto \log {\cal M}  \;\;,
\eeqa
where ${\cal M}$ denotes the moduli-dependent holomorphic mass
of an $N=2$ BPS state.  For the heterotic
$S$-$T$-$U$-$V$ models
under consideration, the tree-level mass ${\cal M}$ is given by
\cite{CLM2}--\cite{NEU}
\beqa
{\cal M}&=& m_2 - i m_1 U \nonumber\\
&+& in_1 T + n_2 (-UT + t V^2) + i b V  \;\;.
\label{holomm}
\eeqa 
Here, $l=(n_1,m_1,n_2,m_2,b)$ denotes the set of integral quantum numbers 
carried by the BPS state. 
The level matching condition for 
a BPS state reads 
\beqa
p^2_L - p^2_R = 2 n^Tm + \frac{1}{2t} b^2  \;\;,
\label{plpr}
\eeqa
where 
\beqa
p^2_R = \frac{|{\cal M}|^2}{2Y} \;\;,\;\;
Y = {\rm Re}T {\rm Re}U - t ({\rm Re}V)^2 >0 \;.
\eeqa
Let us consider states with $n_i=m_j=0$.  For $t=1$, there are two states
with $p^2_L=2$, $p^2_R=0$, carrying $b = \pm 2$.  These two states
are the additional vector multiplets which enhance the $U(1)_V$
to $SU(2)$ at $V=0$.  On the other hand, for $t=2$, there are no such
states with $p^2_L=2$, $p^2_R=0$, reflecting the fact that in this case
there is no gauge symmetry enhancement of the $U(1)_V$ at $V=0$.

Note that the Narain lattice $L$ of signature $(3,2)$ associated with
(\ref{plpr}) is given by 
$L=\Lambda\oplus U(-1)$, 
where $U(-1)$ denotes the hyperbolic plane
$\left(\begin{array}{cc}0&-1\\-1&0\end{array}\right)$, and where
$\Lambda=U(-1)\oplus \\ <2t>=
\left(\begin{array}{ccc}0&-1&0\\-1&0&0\\0&0&2t\end{array}\right)$
in a basis which we will denote by
$(f_2,f_{-2},f_3)$ \cite{gritnik3}.  
 In the basis  
$j_1=f_{-2} + f_2, j_3 =f_{-2},j_4=2 f_2 + 2 f_{-2} -f_3$, 
$L$ is equivalent to the matrix 
\beqa
I=
\left( \begin{array}{ccc} -2&-1&-4\\-1&0&-2\\-4&-2&-8+2t\end{array}\right) 
\;\;.
\label{intersm}
\eeqa
Inspection of (\ref{cubicf}) and (\ref{cubicx4}) shows that
the matrix $-I$ is nothing but the intersection matrix of the $K3$ fibre
of the Calabi--Yau threefolds (\ref{cubicf}) and (\ref{cubicx4})
for $t=1$ and $t=2$, respectively.

Of special relevance to the computation of perturbative
corrections to $F_1$ are those BPS states, whose tree-level mass vanishes
at certain surfaces in the classical
moduli space.  These surfaces are surfaces in
the Siegel modular threefold ${\cal A}_t = \Gamma_t
\backslash {\cal H}_2$.  Here, $\Gamma_t$ denotes 
the target-space duality group, which is a 
paramodular group, $\Gamma_t \subset Sp_4({\bf Q})$ \cite{gritnik3}.
For $t=1$, $\Gamma_1 = Sp_4({\bf Z})$.
The condition ${\cal M}=0$ is the
condition  for a rational quadratic 
divisor 
${\sc H}_l$ \cite{gritnik3}
of discriminant
\beqa
{\rm D}(l)= 2t(p^2_L - p^2_R) = 
4tm_1n_1+4tn_2m_2 + b^2.
\eeqa
The divisors ${\sc H}_l$ with discriminant ${\rm D}(l)$ determine the 
Humbert surface $H_{\rm D}$ in the Siegel modular threefold ${\cal A}_t$
\cite{gritnik3}.  Each Humbert surface $H_{\rm D}$ can be represented by a
linear relation in $T$, $U$ and $V$ \cite{gritnik3}.  
The Humbert surface $H_{\rm 1}$
corresponds to the surface $V=0$.  States becoming massless at
$V=0$ lay on the orbit ${\rm D}(l)=1$, that is, on the orbit $n^Tm=0, 
b^2=1$.  The Humbert surface $H_{\rm 4t}$, on the other hand, corresponds
to the surface $T=U$.  The states becoming
massless on $H_{\rm 4t}$ carry
quantum numbers $n^Tm=1, b=0$.  They carry $p^2_L=2,p^2_R=0$,
and hence, they correspond to the additional vector multiplets
describing the enhancement of $U(1)_T \times U(1)_U$ \cite{CLM2}.

\section{The heterotic perturbative Wilsonian 
gravitational coupling $F_1$}

The heterotic perturbative Wilsonian gravitational coupling
$F_1$ for an $S$-$T$-$U$-$V$ model is, up to ambiguities linear in
$T$, $U$ and $V$, 
 given by \cite{K2,CCL}
\beqa
&&F_1 = 24 S  \label{f1het} \\
&&- \frac{2}{\pi} 
\sum_{k,l,b \in {\bf Z}\atop (k,l,b)>0} 
d(4tkl-b^2) Li_1(e^{-2\pi(kT + lU + bV)}), \nonumber
\eeqa
where the condition
$(k,l,b)>0$ means again that: either $k > 0, l , b \in {\bf Z}$ or 
$k=0,l>0, b \in {\bf Z}$ 
or $k=l=0, b > 0$.  The
expansion coefficients $d(4tN)$ ($N = kl - \frac{b^2}{4t}$)
are determined in terms of \cite{CCL}
\beqa 
F(\tau) &=&  \sum_{N \in {\bf Z}, {\bf Z} + \frac{4t-1}{4t}} c(4tN) q^N\;\;,
\nonumber\\
E_2 F(\tau) &=& 
\sum_{N \in {\bf Z}, {\bf Z} + \frac{4t-1}{4t}} d(4tN) q^N\;\;.
\label{e2f}
\eeqa
Here, the $c(4tN)$ denote the expansion coefficients occuring
in the prepotential (\ref{prepstuv}).
The quantity $E_2 F(\tau)$ should be related to the gravitational threshold
corrections, as follows \cite{MHE,CCL},
\beqa
{\tilde I}_{3,2} = \int_{\cal F} \frac{d^2 \tau}{\tau_2} \Big[
Z_{3,2} F(\tau)[E_2 - \frac{3}{\pi \tau_2}] - d(0) \Big] \;\;.
\eeqa
The Wilsonian coupling $F_1$ should, on the other hand, also be expressible
in terms of target-space duality modular forms, namely 
Siegel modular forms on ${\cal A}_t = \Gamma_t \backslash {\cal H}_2$.
Below, we will rewrite $F_1$ in terms of Siegel modular forms.

\subsection{The $S$-$T$-$U$-$V$ models with $(N_V,N_H)=(5,215-12n)$}

For this class of models, the target space duality group is
$\Gamma_1 = Sp_4({\bf Z})$.  The relevant Humbert surfaces are
$H_{\rm 1}$ and $H_{\rm 4}$.  The Siegel modular forms which vanish 
on $H_{\rm 1}$ and $H_{\rm 4}$ are given as follows \cite{GN2}.
The Siegel modular form which vanishes on the $T=U$ locus and has
modular weight $0$ is given by $\frac{{\cal C}^2_{30}}
{{\cal C}_{12}^5}$.
It can be shown that, as $V \rightarrow 0$,
\beqa
 \frac{{\cal C}^2_{30}}{{\cal C}_{12}^5} \rightarrow (j(T) - j(U))^2\;\;,
\eeqa
up to a normalization constant.
On the other hand, 
the Siegel modular form which vanishes on the $V=0$ locus and has
modular weight $0$ is given by $\frac{{\cal C}_5}
{{\cal C}_{12}^{5/12}}$.
It can be shown that, as $V \rightarrow 0$, 
\beqa
{\cal C}_5 \rightarrow V \left(\eta^{24}(T) \eta^{24}(U) \right)^{\frac{1}{2}}
\;\;,
\eeqa
up to a proportionality constant.   
Here, the Siegel form ${\cal C}_{12}$ is a modular form of weight 12 which
generalises $\eta^{24}(T) 
\eta^{24}(U)$,
that is
\beqa
{\cal C}_{12} \rightarrow \eta^{24}(T) \eta^{24}(U)
\eeqa 
as $V \rightarrow 0$.

The Siegel modular forms ${\cal C}_5$ and 
${\cal C}_{30}$ have infinite product expansions,
as follows \cite{GN2},
\begin{eqnarray}
{\cal C}_{5}&=&(qr^{-1}s)^{1/2}\prod_{k,l,b\in {\bf Z} \atop (k,l,b)>0}
(1-q^kr^bs^l)^{f(4kl-b^2)} \;\;, \nonumber\\
{\cal C}_{30}&=&(q^3r^{-1}s^3)^{1/2}(q-s) \times \label{prodexp}\\
&&\prod_{k,l,b\in {\bf Z} \atop (k,l,b)>0}
(1-q^kr^bs^l)^{f'_2(4kl-b^2)} \;\;, \nonumber
\end{eqnarray}
where $q=\exp{(-2 \pi T)}$, 
$s=\exp{(-2 \pi U)}$ and 
$r=\exp{(-2 \pi V)}$.
The coefficients $f(4kl-b^2)$ and $f'_2(4kl-b^2)$
are defined as follows \cite{GN2}.  Consider the
expansion of the weak Jacobi form of weight 0 and index 1
\beqa
\phi_{0,1}&=&\frac{\phi_{12,1}}{\eta^{24}}
=\sum_{m\ge 0}\sum_{l\in {\bf Z}}
f(4m -l^2)q^mr^l \nonumber\\
&=& (r + 10 +  r^{-1}) \nonumber\\
&+& q (10 r^{-2} - 64 r^{-1} + 108 - 64 r + 10 r^2) 
\nonumber\\
&+& q^2(\dots)
 \;\;,
\eeqa
where the sum over $l$ is restricted to $4n-l^2\geq -1$.  Then,
$f(N)=f(4m-l^2)$ if $N=4m-l^2\geq-1$, and $f(N)=0$ otherwise.
The coefficients $f_2'(N)$ are then given by
$f'_2(N)=8f(4N)+(2\left (\frac{-N}{2}\right )-3)f(N)+f(\frac{N}{4})$.
Here, $(\frac{D}{2})=1,-1,0$ depending on whether $D\equiv 1 \,
{\rm mod} \, 8, 5\, {\rm mod}\, 8,0\,{\rm mod}\, 2$.

Then, in analogy to the perturbative Wilsonian gravitational
coupling $F_1$ for the $S$-$T$-$U$ model \cite{C,CCLMR}
\beqa
F_1 &=& 24 S_{\rm inv} - \frac{b_{\rm grav}}{24 \pi} \log \eta^{24}(T) 
\eta^{24}(U) \nonumber\\
&+& \frac{2}{\pi} \log(j(T) - j(U)) \;\;,
\eeqa
the perturbative Wilsonian gravitational coupling 
for an $S$-$T$-$U$-$V$ model should now be given by (in the 
chamber $T> U>2V$) \cite{CCL}
\beqa
F_1 &=& 24 S_{\rm inv} - \frac{b_{\rm grav}}{24 \pi} \log {\cal C}_{12}
+ \frac{1}{\pi} \log \frac{{\cal C}^2_{30}}{{\cal C}_{12}^5} \nonumber\\
&-& \frac{1}{2 \pi} (N_H'-N_V') \log \left(
\frac{{\cal C}_5}{{\cal C}_{12}^{5/12}}\right)^2 \;\;.
\label{f1stuv}
\eeqa
Here, $N_V'=2$ and $N_H'=12n + 32$ denote the vector and the hyper multiplets
which become massless at the $V=0$ locus.

The invariant dilaton $S_{\rm inv}$ is given by \cite{WKLL}
\beqa
S_{\rm inv} &=& {\tilde S} + \frac{1}{10} L \;\;, \nonumber\\
{\tilde S} &=& S - \frac{4}{10}(\partial_T \partial_U - 
\frac{1}{4} \partial_V^2) h \;\;,
\eeqa
where the role of the quantity $L$ is to render $S_{\rm inv}$ free
of singularities.
It can be shown \cite{CCL} that 
\beqa
{\tilde S} &=& S + \frac{1}{5 \pi} \log \frac{{\cal C}^2_{30}}
{{\cal C}_{12}^5} \\
&-& \frac{3}{10 \pi} (2 + n) \log \left(
\frac{{\cal C}_5}{{\cal C}_{12}^{5/12}}\right)^2 
+ regular \nonumber
\eeqa
and, hence,
\beqa
L=-\frac{2}{\pi} \log \frac{{\cal C}^2_{30}}{{\cal C}_{12}^5} 
+ \frac{3}{ \pi} (2 + n) \log \left(\frac{{\cal C}_5}
{{\cal C}_{12}^{5/12}}\right)^2 \;\;.
\eeqa
It follows that the Wilsonian gravitational coupling (\ref{f1stuv})
can be rewritten as
\beqa
F_1 &=& 24 {\tilde S} \nonumber\\
&-& \frac{1}{\pi} \Big[
\frac{19}{5} \log {\cal C}^2_{30} +  
  \frac{3}{5}(1-2n) 
 \log {\cal C}_5^2 \Big] \;\;. \label{f1tildestuv}
\eeqa
Note that the $\log {\cal C}_{12}$ terms have completely canceled out.

Comparison of (\ref{f1het}) and (\ref{f1tildestuv}) 
\beqa
F_1 &=& 24 S - \frac{2}{\pi} \sum_{(k,l,b)>0} 
d_n(4kl-b^2) Li_1 \label{cons} \\
&=& 24 {\tilde S} - \frac{1}{\pi} \Big[
\frac{19}{5} \log {\cal C}^2_{30}
+ \frac{3}{5} (1 - 2 n)  \log {\cal C}_5^2 \Big] 
\nonumber
\eeqa
gives a highly non-trivial consistency check on (\ref{e2f})
(here, we have ignored the issue of ambiguities linear in $T$, $U$ and
$V$).
It yields, using the product expansions for ${\cal C}_5$
and ${\cal C}_{30}$ given in (\ref{prodexp}), that
\beqa
d_n(N) &=& -\frac{6}{5} N  c_n(N) - \frac{19}{5}
f_2'(N) \nonumber\\
&-& \frac{3}{5}(1-2n)f(N) \;\;,
\label{cff}
\eeqa
where $N = 4 kl - b^2 \in 4 {\bf Z}, 4{\bf Z} + 3$.
In order to show that (\ref{cff}) really holds, consider
introducing the hatted quantities \cite{K2,CCL}
\beqa
{\hat Z}&=&\frac{1}{72}\frac{(E_4^2 \widehat{E_{4,1}}-
E_6\widehat{E_{6,1}})}{\Delta} \;\;,
\nonumber\\
J_C &=& \frac{2 E_6 \widehat{ E_{6,1}}}{\Delta} + 81 {\hat Z} \;\;,
\eeqa
as well as 
\beqa
{\tilde Z}(\tau) &=& {\hat Z}(4 \tau) 
= 2 \sum_{N \in 4 {\bf Z} \, or \,
4{\bf Z} + 3} f(N) q^N , \\ 
{\tilde J_C} (\tau) &=& J_C(4 \tau) = \sum_{N \in 4{\bf Z}, 4{\bf Z}
+ 3} c_J (N) q^{N}. 
\eeqa
Then, it can be verified that
\beqa
f_2'(N) =  \frac{1}{2} c_J (N) + 6 f(N) \;\;.
\label{f2cjf}
\eeqa
It can also be shown that, for $m=4,6$ \cite{K2}
\beqa
\Theta_q E_m &=& \frac{m}{12} \left(E_2 E_m - E_{m+2} \right) \;\;,
\nonumber\\
\Theta_q {\hat E}_{m,1} &=& 
\frac{2m-1}{24} \left(E_2 {\hat E}_{m,1} - 
{\hat E}_{m+2,1} \right) \;\;, \nonumber\\
\Theta_q {\tilde E}_{m,1} &=& 
\frac{2m-1}{6} \left({\tilde E}_2 {\tilde E}_{m,1} - 
{\tilde E}_{m+2,1} \right) \;\;, 
\label{tq}
\eeqa
where 
\beqa
{\tilde E}_2(\tau) = E_2 (4 \tau) \;\;,
{\tilde E}_{m,1}(\tau) = {\hat E}_{m,1} (4 \tau) \;\;,
\eeqa
and
where $\Theta_q=q \frac{d}{dq}$.  Then, by using (\ref{f2cjf}) as
well as (\ref{tq}), 
it can be shown that
(\ref{cff}) indeed holds \cite{CCL}.

\subsection{The $S$-$T$-$U$-$V$ models with $(N_V,N_H)=(5,149)$}

For this class of models, 
the target-space duality group is $\Gamma_2$ 
\cite{gritnik3}.  Thus,
the appropriate Siegel modular forms for this class of models
will be different from the ones considered in the previous
subsection.
Of relevance for the following are going to be the Siegel modular
forms with divisors $H_1$ and $H_8$, denoted by $\Delta_2$ and
$\Psi_{12}^{(2)}$ in \cite{gritnik3}, with modular weight 2 and
12, respectively.  Both $\Delta_2$ and $\Psi_{12}^{(2)}$ enjoy
infinite product expansions, as follows \cite{gritnik3},
\beqa
\Delta_2 &=&q^{1/4}r^{-1/2}s^{1/2} \times \nonumber\\
&&\prod_{k,l,b\in {\bf Z} \atop (k,l,b)>0}
(1-q^kr^bs^l)^{f_2(8kl-b^2)} \;\;, \nonumber\\
\Psi_{12}^{(2)}&=&q\prod_{k,l,b\in {\bf Z} \atop (k,l,b)>0}
(1-q^kr^bs^l)^{c_2(8kl-b^2)} \;\;, 
\label{prodexp2}
\eeqa
where $q=\exp{(-2 \pi T)}$, 
$s=\exp{(-2 \pi U)}$ and 
$r=\exp{(-2 \pi V)}$.
The expansion coefficients $f_2(8kl-b^2)$ are the expansion
coefficients of the following weak Jacobi form of weight 0 and index 2
\cite{gritnik3}
\beqa
\phi_{0,2} &=& \frac{\phi_{2,2}}{\eta^4} 
=\sum_{m\ge 0}\sum_{l\in {\bf Z}}
f_2(8m-l^2)q^m r^l  \nonumber\\
&=& ( r + 4 + r^{-1}) \\
&+& q(r^{\pm 3}-8r^{\pm 2}-r^{\pm 1} + 16)
+ q^2(\dots),\nonumber
\eeqa
whereas the expansion coefficients $c_2(8kl-b^2)$ are the expansion
coefficients of the nearly holomorphic Jacobi form of weight 0 and
index 2 \cite{gritnik3}
\beqa
\Psi_{0,2}&=& \frac{E_{12,2}}{\eta^{24}} 
=\sum_{m\ge -1}\sum_{l\in {\bf Z}}
c_2(8m-l^2)q^m r^l  \nonumber\\
&=& q^{-1} + 24 + q(\dots)
\;\;.
\eeqa
Finally, we will denote the analogue of ${\cal C}_{12}$ by $X_{12}$.  The
Siegel form $X_{12}$ should be such that
it doesn't vanish in the interior of moduli space, 
and that it reduces to $\eta^{24}(T)\eta^{24}(U)$ in the
limit $V \rightarrow 0$.  Below, we will see that $X_{12}$ will 
in the end drop out of the relevant expressions.

Thus, the
Siegel modular form which vanishes on the $T=U$ locus and has
modular weight $0$ is given by $\frac{\Psi_{12}^{(2)}}
{X_{12}}$.  As  $V \rightarrow 0$,
\beqa
 \frac{\Psi_{12}^{(2)}}{X_{12}} \rightarrow T-U\;\;,
\eeqa
up to a normalization constant.
On the other hand, 
the Siegel modular form which vanishes on the $V=0$ locus and has
modular weight $0$ is given by $\frac{\Delta_2^6}
{X_{12}}$.  As $V \rightarrow 0$, 
\beqa
\frac{\Delta_2^6}
{X_{12}} \rightarrow V^6 
\;\;,
\eeqa
up to a proportionality constant.

The perturbative Wilsonian gravitational
coupling should then, in analogy to (\ref{f1stuv}), be given by (in the 
chamber $T> U > 2V$)
\beqa
F_1 = 24 S_{\rm inv} - \frac{b_{\rm grav}}{24 \pi} \log X_{12}
+ \frac{2}{\pi} \log \frac{\Psi_{12}^{(2)}}{X_{12}} \nonumber\\
- \frac{1}{6 \pi} (N_H'-N_V') \log 
\frac{\Delta_2^6}{X_{12}}\;\;.
\eeqa
Using that $ b_{\rm grav}=336$, $N_V'=0$ and $N_H'=96$, it follows that 
\beqa
F_1 = 24 S_{\rm inv} 
+ \frac{2}{\pi} \log \Psi_{12}^{(2)} - \frac{96}{ \pi}  \log \Delta_2 \;.
\eeqa
Note that, in contrast to (\ref{f1stuv}),
 the dependence on $X_{12}$ already drops out when using 
$S_{\rm inv}$.

The invariant dilaton $S_{\rm inv}$ is given by \cite{WKLL}
\beqa
S_{\rm inv} &=& {\tilde S} + \frac{1}{10} L \;\;, \nonumber\\
{\tilde S} &=& S - \frac{4}{10}(\partial_T \partial_U - 
\frac{1}{8} \partial_V^2) h \;\;,
\eeqa
where the role of the quantity $L$ is to render $S_{\rm inv}$ free
of singularities.
Using (\ref{prepstuv}), it follows that, as $T \rightarrow U$,
\beqa
\partial_T \partial_U h &=& - \frac{1}{\pi} \log(T-U) \nonumber\\
&=&
- \frac{1}{\pi} \log \frac{ \Psi_{12}^{(2)}}{X_{12}} + regular
\eeqa
and that, as $V \rightarrow 0$,
\beqa
- \partial^2_V h &=& - \frac{1}{\pi} c(-1) \log V \nonumber\\
&=& \frac{1}{12 \pi} N_H' 
\log \frac{\Delta_2^6}{X_{12}} + regular \;\;.
\eeqa
Hence, 
\beqa
{\tilde S} &=& S - \frac{4}{10 \pi} \left(- \log \Psi_{12}^{(2)} + 
6 \log \Delta_2 \right) \nonumber\\
&+& regular
\eeqa
and
\beqa
L=-\frac{4}{\pi} \left(\log \Psi^{(2)}_{12} 
- 6 \log \Delta_2\right) \;\;.
\eeqa
It follows that 
\beqa
F_1 = 24 {\tilde S} - \frac{1}{\pi} \Big[
\frac{38}{5} \log \Psi_{12}^{(2)} + 
\frac{192}{5} \log \Delta_2 \Big] \;.
\label{f1stuv2}
\eeqa
Note that $X_{12}$ has again cancelled out.

Comparison of (\ref{f1het}) and (\ref{f1stuv2})
\beqa 
F_1 = 24 S - \frac{2}{\pi} 
\sum_{k,l,b \in {\bf Z}\atop (k,l,b)>0} 
d(8kl-b^2) Li_1 \nonumber\\
= 24 {\tilde S} - \frac{1}{\pi} \Big[
\frac{38}{5} \log \Psi_{12}^{(2)} + 
\frac{192}{5} \log \Delta_2 \Big]
\eeqa
gives again a highly non-trivial consistency check on (\ref{e2f})
(here, we have again ignored the issue of ambiguities linear in $T$, $U$ and
$V$).
It yields, using the product expansions for $\Delta_2$
and $\Psi_{12}^{(2)}$ given in (\ref{prodexp2}), that
\beqa
-2d(N) = \frac{6}{5} N c(N) 
+ \frac{38}{5}c_2(N) + \frac{192}{5}f_2(N),
\label{dcc2f2}
\eeqa
where $N=8kl-b^2 \in 8 {\bf Z} , 8 {\bf Z} + 7$.
This relation can be explicitly checked for some of the coefficients.
Consider, for instance, the cases where $N=-8,-1,0$.  Then, using that
\beqa
d(-8)&=&c(-8)=1 \;\;,\;\; d(-1)=c(-1)=-48,\nonumber\\
d(0)&=&c(0) - 24c(-8) = - 168
\eeqa
as well as
\beqa
f_2(-1) &=& 1 \;\;,\;\; f_2(0)=4 \;\;, \nonumber\\
c_2(-8) &=& 1 \;\;,\;\; c_2(0) =24 \;\;,
\eeqa
it is straightforward to 
check that (\ref{dcc2f2}) indeed
holds.

\section*{Acknowledgement}

We would like to thank I. Bakas, P. Berglund, M. Henningson and S. Stieberger
for useful discussions.  We are grateful to P. Mayr for providing
us with the relevant topological data for the $X^{(4)}$ model.

\end{document}